\documentclass[table, screen, acmsmall]{acmart}

\AtBeginDocument{%
  }

\renewcommand\footnotetextcopyrightpermission[1]{} 

\setcopyright{none}
\copyrightyear{2025}
\acmDOI{}

\usepackage{longtable}
\usepackage[flushleft]{threeparttable}
\usepackage{multirow}
\usepackage{pdflscape}
\usepackage{subcaption}
\definecolor{editCol}{rgb}{0.0, 0.0, 0.0}
\newcommand{\edit}[1]{{\textcolor{editCol}{#1}}}
\usepackage{caption}
\usepackage{array}
\usepackage{colortbl} 
\usepackage{comment}
\usepackage{balance}
\usepackage{booktabs} 
\renewcommand{\arraystretch}{1.5} 

\begin{document}

\title[Online Safety for All: Systematic Review of Youth Online Safety in the Global South]{Online Safety for All: Sociocultural Insights from a Systematic Review of Youth Online Safety in the Global South}


\author{Ozioma C. Oguine}
\orcid{0000-0003-2434-1400}
\affiliation{%
\department{Computer Science and Engineering}
  \institution{University of Notre Dame}
  \city{Notre Dame}
  \state{Indiana}
  \country{USA}}
\email{ooguine@nd.edu}

\author{Oghenemaro Anuyah}
\orcid{0000-0002-5773-8239}
\affiliation{%
\department{Computer Science and Engineering}
  \institution{University of Notre Dame}
  \city{Notre Dame}
  \state{Indiana}
  \country{USA}}
\email{oanuyah@nd.edu}

\author{Zainab Agha}
\orcid{0000-0002-6519-3742}
\affiliation{%
\department{Computer Science}
  \institution{San Francisco State University}
  \city{San Francisco}
  \state{California}
  \country{USA}}
\email{zainabagha@sfsu.edu}

\author{Iris Melgarez}
\orcid{}
\affiliation{%
\department{Computer Science and Engineering}
  \institution{University of Notre Dame}
  \city{Notre Dame}
  \state{Indiana}
  \country{USA}}
\email{Ivmelgarez@gmail.com}

\author{Adriana Alvarado Garcia}
\orcid{0000-0002-4230-3777}
\affiliation{
\institution{IBM Research}
  \city{New York}
\country{United States}
}

\author{Karla Badillo-Urquiola}
\orcid{0000-0002-1165-3619}
\affiliation{%
\department{Computer Science and Engineering}
  \institution{University of Notre Dame}
  \city{Notre Dame}
  \state{Indiana}
  \country{USA}}
\email{kbadillou@nd.edu}

\renewcommand{\shortauthors}{O.C Oguine et al.}

\begin{abstract}
Youth online safety research in HCI has historically centered on perspectives from the Global North, often overlooking the unique particularities and cultural contexts of regions in the Global South. This paper presents a systematic review of 66 youth online safety studies published between 2014 and 2024, specifically focusing on regions in the Global South. Our findings reveal a concentrated research focus in Asian countries and predominance of quantitative methods. We also found limited research on marginalized youth populations and a primary focus on risks related to cyberbullying. Our analysis underscores the critical role of cultural factors in shaping online safety, highlighting the need for educational approaches that integrate social dynamics and awareness. We propose methodological recommendations and a future research agenda that encourages the adoption of situated, culturally sensitive methodologies and youth-centered approaches to researching youth online safety regions in the Global South. This paper advocates for greater inclusivity in youth online safety research, emphasizing the importance of addressing varied sociocultural contexts to better understand and meet the online safety needs of youth in the Global South.

\end{abstract}



\maketitle

\section{Introduction}
 A recent Microsoft Global Online Safety Survey disclosed that 69\% of adolescents across 14 countries encountered online risks in the past year \cite{NewMicro74:online}. Online safety is a critical and global concern that has received significant attention from the CSCW \cite{agha2023strike, badillo2021conducting} and broader human-computer interaction (HCI) community \cite{badil2risk2restriction, pinter2017adolescent,wilkinson2022many,freed2023understanding}. While technology provides several benefits to youth such as social support \cite{Oguine2024CHINS, razi2020lets, lavis2020online, tachtler2020supporting} and access to information and education \cite{gogus2019privacy, Reeves2024Jun}, technological advancements, especially in social media, continue to introduce online risks \cite{ali2022understanding}, ranging from cyberbullying, information related risks, sexual solicitations, child sexual abuse material (CSAM) to more recent risks posed by generative AI \cite{razi2020lets, thiel2023generative}. This has led to recent legislative policies such as the Kids Online Safety Act (KOSA) \cite{Sen.Blumenthal2023Dec} in the US and EU Digital Services Act \cite{EUsafetyAct}, along with social media companies introducing parental controls and age verification \cite{TikTok2024Oct}, depicting a "moral panic" around online safety of youth at a global scale \cite{oguineKOSA}. 

Unfortunately, much of the research and legislative efforts concerning online safety for youth are from a Global North perspective \cite{pinter2017adolescent}. Pinter et al. \cite{pinter2017adolescent} conducted a systematic literature review in 2017, highlighting the need to prioritize a globally inclusive approach that considers the diverse cultural, ethnic, and political factors influencing adolescent online safety. Addressing this significant gap is crucial to ensure a more diverse and inclusive youth online safety research. The experiences and challenges faced by adolescents in the Global South\footnote{In this work, the term \textit{``Global South''} refers to a wide range of regions that are known for their rich cultural heritage, which is deeply rooted in diverse traditional practices and unique cultural expressions \cite{Dados2012Feb}}. This covers countries primarily in Latin America, Africa, the Middle East, and parts of Asia \cite{globalsouthcountries}, which are often underrepresented in mainstream HCI research \cite{anuyah2023cultural}, for instance, may differ significantly from those in more developed countries due to diverse cultural norms, political landscapes, and levels of digital literacy. Expanding research to include these varied contexts is essential for developing a more comprehensive and culturally sensitive understanding of online safety for adolescents worldwide. Our work addresses this critical gap by exploring adolescent online safety practices and contributions in the Global South, considering how cultural, ethnic, and political contexts shape these experiences. Specifically, we ask the following research questions:
 
\begin{itemize}
    \item \textbf{RQ1:} How has youth online safety been studied in the Global South?
    \item \textbf{RQ2:} What are the major contributions of the research on youth online safety in the Global South?
    \item \textbf{RQ3:} How can HCI research take a more holistic and global approach to investigating youth online safety?
\end{itemize}

To address these questions, we conducted a systematic review of 66 research articles focused on youth online safety in the Global South. We employed a qualitative approach, using both open and axial coding methods to understand the research scope across four dimensions: \textit{where} the research took place (region of focus), \textit{who} the studies involved, \textit{how} the data was collected and how they contribute to the broader body of knowledge, and \textit{what} key factors or measures are examined. To answer (RQ1 and RQ2), we found that cyberbullying and online victimization were the most frequently studied online risks in the Global South, reflecting widespread concern over their emotional and psychological impact on youth. Privacy concerns and \edit{online sexual risks\footnote{\edit{We define "online sexual risks" as potential dangers associated with engaging in sexual behaviors or with sexual content through the internet (e.g., sexual grooming, online child sexual exploitation and abuse (OCSEA), and child trafficking) \cite{reeves2024May}.}}} were also significant but less frequently addressed, indicating areas where further attention may be warranted. The majority of online safety research was conducted in Asian countries, such as India and China, with comparatively fewer studies in African, Latin American, and other regions, suggesting a regional imbalance in the literature. Research topics were often shaped by the unique cultural norms, education levels, and socioeconomic conditions of each region. For example, studies noted how local values and societal expectations influenced youth experiences of cyberbullying and privacy. In response to these risks, strategies in the Global South predominantly emphasized education and awareness campaigns to improve digital literacy and equip youth with skills to navigate online environments. Parental involvement was also frequently highlighted as a protective factor, with parents playing a crucial role in guiding and monitoring their children's online activities to promote safer behaviors.

To answer RQ3, we examined ways in which HCI research could adopt a more holistic and globally inclusive approach to investigating youth online safety. Our analysis suggests the need for greater methodological diversity, including more qualitative and participatory methods that allow youth to voice their experiences directly. Additionally, we found that addressing youth online safety in Global South regions requires culturally responsive frameworks that account for regional socio-economic factors, community norms, and local practices. Encouraging cross-regional collaborations between researchers in the Global South and Global North can also enrich HCI research by integrating diverse perspectives, thus fostering more comprehensive and contextually relevant solutions for youth online safety across various digital environments.

Our research makes theoretical contributions to the broader CSCW community and the field of youth online safety by offering a nuanced understanding of the research landscape, trends, and practices that shape youth online safety in the Global South. This paper, therefore, makes the following novel contributions:

\begin{itemize} 
    \item A systematic literature review of the past decade of HCI research on online safety in the Global South, offering a deeper understanding of the influence of socio-cultural factors on youth online safety. 
    \item A comprehensive synthesis of the unique contributions and findings of youth online safety research across different regions of the Global South, highlighting both common and region-specific online risks and protective strategies. 
    \item Practical implications and recommendations for advancing HCI research on youth online safety, encouraging more culturally informed and globally inclusive perspectives. 
\end{itemize}

In conclusion, this review not only broadens the scope of HCI research by integrating diverse sociocultural perspectives from the global south but also highlights critical areas for future research. By prioritizing youth voices, fostering international collaborations, and integrating culturally responsive approaches to research, future studies can more effectively address the unique needs of youth in the Global South, promoting safer and more inclusive digital environments for young people worldwide.

\section{Background}
In this section, we first examine existing approaches to youth online safety within HCI, synthesizing studies focused primarily on adolescents' interactions with digital platforms in Western contexts. We then transition to exploring the emerging HCI research focused on adolescents' online safety in the Global South.

\subsection{Existing Approaches to Youth Online Safety in HCI} 
Research on adolescent online safety in HCI has evolved over time, with a focus on the myriad online risks faced by youth, including cyberbullying, sexting, and exposure to harmful content across digital platforms such as social media \cite{razi2020lets, badillo2017abandoned, schoenebeck2023online, callaghan2015exploring}, social virtual reality \cite{oguine2023you, malone}, and more recently, generative AI \cite{noraAI, genAI}. Studies have consistently highlighted the vulnerabilities adolescents experience when navigating these platforms. For example, Razi et al. \cite{razi2020lets} explored how adolescents aged 12–17 encountered sexual risks when seeking social support online, while Wisniewski et al. \cite{pam2016deardiary} found that teens regularly engage in risky online activities such as sexual solicitations and cyberbullying. These findings underline the ongoing need for safety interventions tailored to adolescents' online experiences. Acknowledging the dual nature of online technologies--providing both opportunities and risks for adolescents--scholars have proposed a range of protective strategies. Initially, much of the focus was on parental mediation, emphasizing the role of parents and caregivers in monitoring and restricting teens' online interactions to ensure a safer environment \cite{marjie2015, minsam2015}. However, recent studies have uncovered several limitations to this approach. For instance, stringent monitoring is often perceived by adolescents as invasive and privacy-violating, which can hinder their developmental growth and limit their opportunities for self-regulation in online environments \cite{wisniewski2018theprivacy, pam2017parents, agha2023strike}.

In response to these concerns, the focus of HCI research has shifted towards more autonomy-supportive and resilience-based interventions that empower teens to navigate online risks independently while fostering their development. Agha et al. \cite{agha2023strike}, for instance, investigated adolescents’ preferences for nudges--design-centric interventions that allow them to take control of their online safety. This preference for self-regulation highlights a growing recognition of the tension between parental control and the autonomy that teens seek in managing their own online safety. Similarly, Badillo et al. \cite{badillo2020beyond} emphasized the importance of strength-based approaches that equip adolescents with the tools to handle online risks proactively, thus moving beyond parental supervision and towards a more balanced and teen-centric approach to safety interventions.

Despite these advancements, much of the existing research is focused on adolescents from Western contexts, as noted by Pinter et al. \cite{pinter2017adolescent} in their systematic review of online safety studies. This review revealed that a significant portion of the research is centered on the United States and Europe, with limited representation from other regions, particularly the Global South. The review also found that studies in Western contexts often employ different approaches to understanding adolescent online safety, which may not be directly transferable to non-Western settings. Since the review by Pinter et al. \cite{pinter2017adolescent} , the landscape of social media and digital platforms has shifted significantly, further amplifying the need to explore how online safety risks and strategies are evolving across diverse global contexts.

\subsection{Investigating Youth Digital Experiences in the Global South}

Building on Pinter et al. \cite{pinter2017adolescent}’s observations, our review extends the discussion to the Global South, where the sociocultural and technological landscape differs significantly from Western settings \cite{anuyah2023cultural}. HCI research focusing on adolescents in the Global South is growing but remains limited, often lacking a dedicated exploration of online safety risks. Population in the Global South represents roughly 85\% of the world's population \cite{GlobalAsian}, making it essential to examine how these factors shape adolescents' online experiences and vulnerabilities.

A recurring theme in the Global South is the digital divide, which affects not only access to technology but also digital literacy and online participation. Studies such as Yudan-Ma et al. \cite{ma2018effects} highlight how playful technological interventions, like audio augmentation in education, can address educational disparities. Similarly, Tachtler et al. \cite{tachtler2020supporting} focus on mental health care for unaccompanied migrant youth, demonstrating how digital tools can be used to provide critical support to vulnerable adolescents. These studies show the diverse contexts in which youth interact with technology, yet they often do not directly address the specific online safety risks that these adolescents face. Moreover, socio-cultural factors in the Global South further complicate the landscape of adolescent online safety. Family structures, social norms, and power dynamics play important roles in shaping how adolescents navigate online spaces. For example, in certain regions, societal taboos around discussing sexual health or gender-related issues create additional risks for teens who seek information or support online. Research by Karusala et al. \cite{karusala2021that} on adolescents living with HIV in rural Kenya demonstrates how mobile platforms can be used to foster peer support networks while navigating significant privacy concerns. The study illustrates how the design of digital interventions must consider cultural sensitivities to avoid exacerbating vulnerabilities.


Despite the growing interest in HCI research in the Global South, studies focusing on adolescent online safety remain scarce. While existing research provides insights into the broader interactions between culture and technology, it often lacks a focused exploration of how online risks are experienced and mitigated by adolescents in these regions. Studies that do address these issues tend to be fragmented, addressing only specific risks or interventions without taking into account the broader socio-political and economic factors that shape adolescents’ digital lives. Our systematic review addresses this gap by offering a comprehensive analysis of the current state of research on adolescent online safety in the Global South. We investigate how online safety risks are studied in these regions and assess the contributions and limitations of existing research. This analysis is critical for developing context-specific interventions that cater to the unique needs and challenges of adolescents in the Global South. By doing so, we aim to contribute to a more globally inclusive approach to youth online safety in HCI, one that moves beyond Global North-centric frameworks and engages with the diverse realities of adolescents worldwide.

\section{Terminology and Positionality}
In this section, we establish the foundational terms and perspectives that guide our research. By defining key concepts and acknowledging our positionalities, we aim to provide clarity and context to our work. 

\subsection{Seeking a Common Definition: Online Safety and Risks}
In this paper, we frequently use the terms ``online safety'' and "online risks" as they are central to our aim of understanding youth online safety research efforts on a global scale. Within the broader literature on online safety in HCI, several different terms have been used for online safety (e.g., ``well-being'', "trust and safety", "privacy and safety") and online risks (e.g., "harms", "threats", "vulnerabilities", "exploitation") that refer to intersecting ideas on protecting users from harmful encounters online \cite{Aghatricky, Liu2019, Ali2020Dec, pam2021heidi, razi2020lets}. Using these adjacent terminologies, online safety and risks have been defined in several ways, such as through developing frameworks \cite{tsai2016understanding}, operationalizing various forms of risks \cite{wisniewski2013grand}, or conceptualizing protective strategies for safety \cite{boehmer2015determinants}. A commonality amongst most of these definitions is that they define safety as the absence of risk and harm. Livingstone \cite{livingstone2013online} has been one of the first to reflect on the complex nature of online risks and harm, noting that risks may not always equate to harm and that an understanding of which risks cause harm and why may represent online risks more accurately. In their work, it is also emphasized to consider various aspects of risks such as the environment and the reactions, and to ask important questions such as \textit{"Is it harmful? Who is vulnerable? What can be done?"}. Building upon this work, Wisniewski et al. \cite{wisniewski2013grand} provided a framework that operationalized online safety/risk as a four-stage process, including risky behavior, risk events, responses, and outcomes. Most recently, Agha et al. \cite{Aghatricky} found that youth' consider online risks and safety to be on a spectrum rather than discrete categories, where risky and safe traits can co-exist in subtle ways. Given the nuanced nature of online safety and risks, there is a lack of consensus on defining these terms, and a more holistic understanding is needed to understand global perspectives on this topic. Therefore, in this work, we take a more open-ended approach to understand how youth online safety and risks are operationalized in research across \edit{the Global South}.

\subsection{Distinguishing between Children, Adolescents, and Youth}\label{sec:youthdef}
We align with the United Nations' ``Convention on the Rights of the Child,'' which defines a child as any human being below the age of eighteen \cite{childDefinition}. Yet, this broad definition includes children, adolescents (teens), and youth. Based on the well-known eight stages of psychosocial development, childhood is defined as the stage between birth and adolescence (13-18 years old) \cite{simplyPsych2024}. Understanding that youth are not a homogeneous group is important for ensuring online safety strategies and interventions are developed appropriately for this demographic \cite{parkTeenCentric}. In this paper, rather than using children as an all-encompassing term, we use youth to describe the wide age range from early childhood to the brink of adulthood, a period marked by significant development and vulnerability. This perspective strengthens our commitment to addressing the complex challenges of online safety for global youth, emphasizing the need for robust protective measures and informed, sensitive policies.

\subsection{Addressing the Geopolitical Distinctions between the Global South and Global North}
The term `Global South' covers regions known for their rich cultural heritage, rooted in diverse traditional practices and unique cultural expressions \cite{Mustafa2023Nov, kaggwa2024evaluating}. It primarily includes countries in Latin America, Africa, the Middle East, and parts of Asia \cite{globalsouthcountries}. For specificity, we referred to the United Nations Office for South-South Cooperation (UNOSSC) database, where countries in the Global South are defined economically \cite{globalsouthcountries}. This perspective helped us understand the varied socio-economic and cultural landscapes that shape youth online experiences. In contrast, the `Global North' typically refers to more economically developed and industrialized countries with different cultural and technological contexts. This covers countries in North America, Europe, and Oceania (Australia and New Zealand), as well as Israel, South Korea, and Japan. For our work, this dichotomy is essential for exploring how global disparities influence online safety and risk perceptions among youth, providing a comprehensive view of the challenges and protective measures in varied settings.

\subsection{Positionality}
We are a research team of two PhD students, a community service worker, an industry professional, and two faculty members. Together, we originate from regions in the Global South: Africa (2), Asia (1), and Latin America (3). Each team member is committed to understanding and uplifting vulnerable and marginalized communities, with a specific focus on their navigation of the digital spaces. Our expertise, though varied, is complementary. Some members bring extensive experience in social services, while others have devoted many years to studying youth online safety. This multidisciplinary approach enables us to address the complexities of youth online safety from a comprehensive perspective.

United by a commitment to enhancing the well-being of marginalized and vulnerable groups in a technological society, our research is motivated by a collective vision. We are dedicated to leveraging our diverse insights to create environments where youth can effectively navigate online spaces. Through this perspective, we aim to contribute to the development of online safety measures that are not just effective but are also culturally sensitive and inclusive. This ensures that youths from all backgrounds can navigate the digital world securely and confidently.

\section{Method}
\begin{figure}
     \centering
     \includegraphics[scale=0.6]{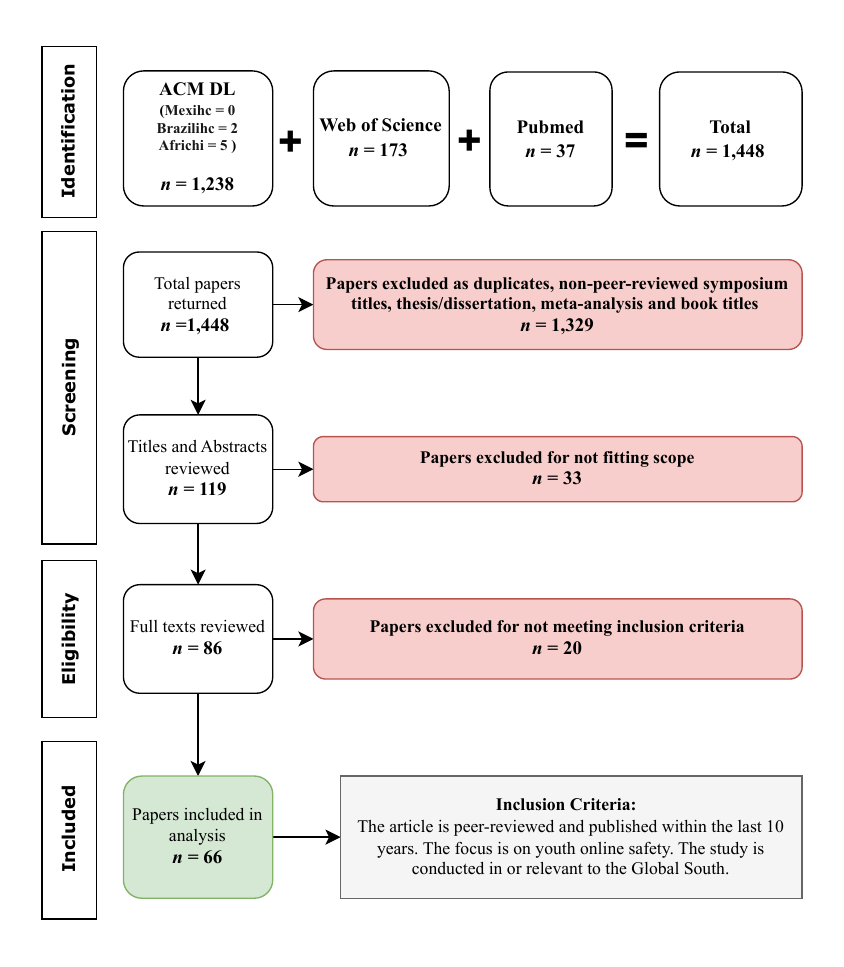}
          \caption{Search, Screening, and Selection Process (Flow Diagram)}
     \label{fig:prisma_flow}
 \end{figure}

To explore the landscape of youth online safety in the Global South, we conducted a systematic literature review, following guidelines adapted from the widely recognized Preferred Reporting Items for Systematic Reviews and Meta-Analyses (PRISMA) framework \cite{page2021prisma, moher2010preferred}. This approach ensures rigor and clarity in our systematic review process. Figure \ref{fig:prisma_flow} illustrates our process, which involved: (1) defining a search strategy, (2) establishing inclusion and exclusion criteria, (3) conducting multiple rounds of screening to select relevant studies, and (4) analyzing the final set of articles. Each of these steps is described in more detail below.

\subsection{Search Strategy: Keywords and Databases}

To capture a broad spectrum of HCI literature on youth online safety, we selected three databases: 1) ACM Digital Library, 2) Web of Science, and 3) PubMed. The ACM Digital Library is a crucial resource for HCI research, as it hosts proceedings from 26 HCI-related conferences sponsored by ACM's SIGCHI (The Special Interest Group on Computer–Human Interaction) \cite{SIGCHI}. The Web of Science provides global coverage of social and information sciences, drawing from multiple publishers such as IEEE, Elsevier (ScienceDirect), and Taylor \& Francis (Tandfonline). PubMed is widely recognized for its focus on health-related literature, and it added a complementary dimension to our search. We specifically targeted HCI proceedings for Global South regions by incorporating papers from conferences like MexIHC (focused on HCI research in Mexico) \footnote{The biannual conference on human-computer interaction is organized by the Mexican Association on Human-Computer Interaction (AMexIHC). The conference proceedings have been published in ACM since the third edition of the conference held in 2010.}, BrazilIHC (HCI research in Brazil) \footnote{The annual conference on human-computer interaction is organized by the Brazilian Symposium on Human Factors in Computing Systems (BrazilIHC). The conference proceedings have been published in ACM since the seventeenth edition of the conference held in 2016.}, and AfriCHI (HCI research in Africa) \footnote{The biannual conference on human-computer interaction is organized by the African Symposium on Human Factors in Computing Systems (AfriCHI). The conference proceedings have been published in ACM since the seventeenth edition of the conference held in 2016.}. While these conferences are part of the ACM, searching them individually yielded different results compared to a general search in the ACM Digital Library. This differentiation highlights the importance of including region-specific conferences. We searched for articles between 2014 to 2024.

\begin{table}[ht]
\caption{Key terms related to online safety, the Global South, and youth.}
\label{tab:keyterms}
\centering
\footnotesize
{\renewcommand{\arraystretch}{1.5}
\begin{tabular}{p{3cm}|p{8cm}}
\hline
\rowcolor[HTML]{EFEFEF}
\textbf{Category}            & \textbf{Related Terms}                                                                                                               \\ \hline
\rowcolor[HTML]{FFFFFF}
\textbf{Online Safety}    & Online safety, digital safety, cybersecurity, internet safety, digital security, cybersafety, online security, internet security, social media safety. \\ \hline
\rowcolor[HTML]{EFEFEF}
\textbf{Global South}    & Global South, developing countries, developing regions, developing world, Africa, Sub-Saharan Africa, South Africa, Latin America, Central America, South America, Southeast Asia, Asia, Asia-Pacific, Caribbean, LMIC, low-income, emerging economies, international development, marginalized, underserved, Angola, Botswana, Democratic Republic of the Congo, Ethiopia, Ghana, Kenya, Nigeria, Rwanda, Tanzania, Uganda, Zambia, Zimbabwe, Algeria, Egypt, Libya, Morocco, Tunisia, Afghanistan, Bangladesh, India, Nepal, Pakistan, Sri Lanka, Cambodia, Indonesia, Laos, Myanmar, Philippines, Thailand, Timor-Leste, Vietnam, Iraq, Jordan, Lebanon, Syria, Yemen, Kyrgyzstan, Tajikistan, Uzbekistan, Argentina, Bolivia, Brazil, Chile, Colombia, Ecuador, Paraguay, Peru, Uruguay, Venezuela, Costa Rica, El Salvador, Guatemala, Honduras, Nicaragua, Panama, Cuba, Dominican Republic, Haiti, Jamaica, Trinidad and Tobago, Fiji, Papua New Guinea, Solomon Islands, Vanuatu. \\ \hline
\rowcolor[HTML]{FFFFFF}
\textbf{Youth} & Youth, teenagers, teens, adolescents, young people, young adults, child, children. \\ \hline
\end{tabular}
}
\end{table}

To create our search query \autoref{tab:Search_string_1}, we used keywords related to online safety, youth, and terms specific to the Global South (See \autoref{tab:keyterms}). To ensure thorough coverage of the Global South, we included the names of all countries classified under this category by the United Nations Office for South-South Cooperation (UNOSSC) \cite{globalsouthcountries}. We also included economic terms such as LMIC (Low- and Middle-Income Countries). We employed Boolean logic, using “OR” to account for alternative terms within each category and “AND” to connect the main keyword groups. This search strategy was applied across the selected databases (covering sections like titles, keywords, and abstracts), yielding a total of (N=1,448) articles: 1,238 from ACM (including \textit{MexIHC, BrazilIHC, and AfriCHI}), 173 from Web of Science, and 37 from PubMed.

    \begin{table}[ht]
        \small
        \centering
        \caption{Full resulting search query.}
        \label{tab:Search_string_1}
        \begin{tabular}{p{12cm}}
            \hline
            \rowcolor[HTML]{FFFFFF}
            \textbf{Query Metadata} \\
            \hline \hline
                \textbf{ Search 1:    Search Query for Online Safety Policies for Youth in the Child Welfare System} \\
                \begin{tabular}{@{}p{12cm}@{}}
              ("online safety" OR "digital safety" OR "cybersecurity" OR "internet safety" OR "digital security" OR "cybersafety" OR "online security" OR "internet security" OR "social media safety")\textbf{ AND }("Global South" OR "developing countries" OR "developing regions" OR "developing world" OR "Africa" OR "Sub-Saharan Africa" OR "South Africa" OR "Latin America" OR "Central America" OR "South America" OR "Southeast Asia" OR "Asia" OR "Asia-Pacific" OR "Caribbean" OR "LMIC" OR "low-income" OR "emerging economies" OR "international development" OR "marginalized" OR "underserved" OR \textbf{[LIST OF GLOBAL SOUTH COUNTRIES, e.g., ‘Algeria,’...]}) \textbf{AND} ("teenagers" OR "teens" OR "adolescents" OR "youth" OR "young people" OR "young adults" OR "Children" OR "Child") \\
                \end{tabular} \\
                \hline
        \end{tabular}
    \end{table}

 \subsection{Selection Criteria and Paper Screening}

Before screening, we eliminated duplicates and non-research articles, narrowing our dataset to 1,448 articles. To ensure relevancy and quality, we applied the following inclusion criteria in selecting papers: First, papers must have been peer reviewed and published in a journal or conference proceeding. Second, the research must have focused on youth aged 19 and younger (see definition in section \ref{sec:youthdef}). Third, the study must have directly addressed online risks, with specific attention to research conducted in or about regions within the Global South. To enhance the diversity of our dataset, we also included papers published in Spanish. Additionally, the research must have contributed to the field of HCI (defined by Wobbrock's 7 types of HCI contributions: empirical, artifact, methodological, theoretical, dataset, surveys, or opinions~\cite{wobbrock}. This last criterion ensures that the selected papers are relevant and contribute meaningfully to HCI.

\edit{While our search strategy was designed to maximize representation across the Global South, we acknowledge that certain regions or countries had limited research output in HCI venues. To address this, we explicitly included region-specific conferences (MexIHC, BrazilIHC, AfriCHI) and incorporated Spanish-language papers to capture research beyond English-language publications. Additionally, we included studies from multidisciplinary databases (e.g., Web of Science and PubMed) to broaden the scope beyond ACM-centric HCI research. By using a comprehensive search query that incorporated both broad terms (e.g., ``Global South'') and country-specific keywords, we ensured that diverse perspectives were included while remaining systematic in our selection process.}

The first author conducted the initial screening, assessing the relevance of each paper based on its title and abstract. If a paper’s title suggested it was relevant to youth online safety, it was passed to the abstract screening stage. After reviewing titles and abstracts, 119 papers remained. The first and fourth authors then conducted a full-text review of the papers, resulting in a final corpus of (N=66) papers for in-depth analysis.

\subsection{Data Analysis Approach}
We employed reflexive thematic analysis \cite{Braun2021Oct} to thoroughly explore the landscape of youth online safety research in the Global South over the past decade. Initially, the first author randomly selected (n=6) $\approx$ 10\% papers from the final dataset of (N=66) articles and carefully reviewed the papers to develop an initial set of dimensions and codes that capture key concepts and ideas prevalent across the studies. Subsequently, the first and fourth authors, \edit{with input and guidance from the last two authors,} applied this preliminary codebook \edit{through multiple coding rounds} on the remaining papers to identify key themes and patterns across the articles, iteratively refining and updating codes as new insights emerged. Articles written in Spanish were also coded by the fourth author, who is fluent in Spanish. \edit{Since the two primary coders worked together closely during the analysis phase, following a consensus-coding approach, calculating inter-rater reliability was not needed \cite{nora_mcdoanld}. Any disagreements between the two coders were resolved in regular discussions and meetings with the last two authors serving in an advisory role.} Our final codebook can be found in Appendix \ref{tab:codebook}. In the next section, we present in detail our findings from the structured analysis.

\section{Findings}
In this section, we present the key findings from our systematic review of youth online safety articles from the Global South. We start by outlining \textit{where} the research was conducted, highlighting geographical trends. We then explore who the studies focused on, including population demographics such as youth and marginalized groups. Next, we examine how the research was conducted, detailing the methods and data collection strategies used. Finally, we analyze what risks were identified and why certain social factors, such as culture and education, were emphasized, as shown in \autoref{tab:codebook}.

\subsection{Where: Most Research Conducted on Asia}
Our analysis revealed a growing body of youth online safety research across the Global South, with studies spanning twenty-seven (27) Global South countries and collaborative efforts from eight (8) Global North nations. Asia emerged as the leading contributor, with Southeast Asia producing the most articles (N=18) \cite{Shin2014Aug, Mamat2023Apr, Arifin2019Jun}, followed by Southern Asia (N=15) \cite{Ghorashi2019Feb, Mustafa2023Nov} and Africa (N=13) \cite{BlessingNamibia, Chipangura2022COVID, Farhangpour2019Mar}. Eastern Asia also contributed significantly (N=11) \cite{Wang2022May, Chen2023}, as did South America (N=8) \cite{Silva2017Privacy, Rivera2016}. Certain countries, such as China (11 articles) \cite{Leung2023Autism}, India (9 articles) \cite{Gan2022}, Malaysia (6 articles) \cite{MohdFadhli2022May}, and South Africa (4 articles) \cite{Chipangura2022COVID}, stood out as potential research hubs, although our analysis did not uncover any specific factors explaining the prominence of these countries.

Interestingly, While many studies were regionally confined \cite{Gan2022, Chipangura2022COVID, Leung2023Autism}, a few adopted cross-regional perspectives \cite{wachs2020routine, Pangrazio2020Jun}, reflecting emerging collaborations between Global North and South. For example, comparative studies conducted by Wachs et al.\cite{wachs2020routine} to understand the routine of cybergrooming victimization among adolescents reported the collection of cross-sectional survey data that spans countries in Southern Asia, Europe, and North America, and another covers Western Asia, Europe, and North America \cite{WachsHITem}, indicating a trend toward international collaboration. Some countries, like Nepal, Iran, Bangladesh, The Gambia, and Chile, had fewer studies (N=1) per country, highlighting the need for a broader geographic focus. Notably, our analysis found that most research from the Global South (N=51) were conducted by scholars affiliated with institutions in the region, demonstrating the interests and dedication of local researchers to advancing the field of youth online safety. While cross-regional studies (N=15) that compared youth online safety across the Global South and Global North regions were typically collaborative efforts between researchers from both regions, demonstrating the importance of global partnerships in advancing youth online safety research.\\

\begin{table}[ht]
    \caption{Geographic Landscape on Youth Online Safety (2014-2024) in the Global South}
    \label{tab:geographical_trend}
    \centering
    \footnotesize
    {\renewcommand{\arraystretch}{1.5}
    \begin{tabular}{p{3cm}|p{7cm}|p{2cm}}
        \hline
        \rowcolor[HTML]{EFEFEF}
        \textbf{Global South Region} & \textbf{Countries (number of articles)} & \textbf{No. of countries} \\ \hline
        \rowcolor[HTML]{FFFFFF}
        Africa & Eritrea (1), Kenya (2), Namibia (3), Nigeria (2), South Africa (4), The Gambia (1) & 6 \\ \hline
        \rowcolor[HTML]{EFEFEF}
        Eastern Asia & China (11) & 1 \\ \hline
        \rowcolor[HTML]{FFFFFF}
        South America & Brazil (2), Chile (1), Colombia (2), Mexico (2) Uruguay (1) & 5 \\ \hline
        \rowcolor[HTML]{FFFFFF}
        South-East Asia & Cambodia(1), Indonesia (2), Malaysia (6), Philippines (1), Taiwan (1), Thailand (5), Vietnam (2) & 7 \\ \hline
        \rowcolor[HTML]{EFEFEF}
        Southern Asia & Bangladesh (1), India (9), Iran (1), Nepal (1), Pakistan (2), Sri Lanka (1) & 6 \\ \hline
        \rowcolor[HTML]{EFEFEF}
        Western Asia & Saudi Arabia (3), Syria (1) & 2 \\ \hline
        \rowcolor[HTML]{FFFFFF}
        Global North Collaborations & Australia (1), Cyprus (1), Germany (1), Greece (1), Singapore (1), South Korea (2), Spain (1), USA (1) & 8 \\ \hline
    \end{tabular}
    }
\end{table}

\subsection{Who: Limited Studies Focusing on Marginalized Youth Populations}
Our analysis revealed that while our dataset focused on articles about the online safety of youth, the investigations were conducted with several different types of stakeholders, such as youth (N=54), parents (N=10), teachers (N=1), and counselors (N=1). 

Most studies investigated online safety from youth perspectives (N=54), using self-reported surveys to understand their online experiences \cite{Vieira2022May, MohdFadhli2022May}. Of these studies, most (N=50) explored gendered perspectives by comparing the online experiences of male and female youth. For instance, Fadhli et al. \cite{MohdFadhli2022May} investigated the connection between cyberbullying and suicidal behaviors among male and female adolescents during the COVID-19 pandemic using a questionnaire. Most of these studies, however, focused on investigating gender experiences through a binary male and female lens, with little studies on non-binary or diverse gender identities. We only found two papers on LGBTQIA+ youth \cite{Hendricks2020Apr} and youth with disabilities \cite{Leung2023Autism}. 

While analyzing these articles, we found that many regions lacked demographic data on LGBTQIA+ youth and youth with disabilities, often due to data collection practices that did not account for these identities or associated online risks \cite{Hendricks2020Apr}. Cultural barriers also contributed to this, as discussing LGBTQIA+ identities was socially sensitive or legally restricted in some Global South countries. Such barriers led to reluctance in openly addressing these topics in research, further limiting the inclusion of marginalized groups in youth online safety studies \cite{Leung2023Autism, Hendricks2020Apr}.

Beyond youth perspectives, several articles studied youth online safety through the lens of other stakeholders (N=12). Most including the perspectives of parents (N=10), either independently (N=3) \cite{Reginasari2021} or alongside youth (N=7) \cite{parentdyads, Achuthan2022Nov}, recognizing the influential role parents played in youth online safety. Some studies also engaged other stakeholders, such as teachers (N=1) and counselors (N=1), illustrating a broader interest in understanding the ecosystem surrounding youth online safety. These studies focused on exploring how different stakeholders and contexts contributed to shaping youth online safety practices across the Global South.

\subsection{\edit{How: Quantitative and Empirical Approaches Predominantly Used to Investigate Youth Online Safety}}
Our analysis revealed a significant predominance of articles using quantitative methods (N=66), particularly survey-based studies (N=43). Most studies adopted a cross-sectional design \cite{Vieira2022May, MohdFadhli2022May, Leung2023Autism}, using standardized questionnaires to explore key trends and relationships within the data. These studies typically used statistical analyses to establish correlations and inform findings. Notably, only one article used a longitudinal survey approach, which offered valuable insights into how online safety behaviors evolve over time \cite{Kunwar2024Mar}. This method, though rare in the literature, proved especially useful in studying the evolving patterns of risks among youth. Kunwar et al. \cite{Kunwar2024Mar} examined cyberbullying and cyber-victimization patterns of youth in Nepal over a 30-day period, highlighting shifts in both victimization rates and online behavior patterns.

In contrast, only (N=17) articles employed qualitative methods \cite{Leung2023Autism, Tech_Addiction}, which delve into deeper contextual understanding and participant experiences. These studies primarily used interviews (N=8), focus groups (N=1), case studies (N=1), and user studies (N=2) to delve into individual experiences and the complexities of online interactions. Only a few studies (N=3) employed mixed methods, combining qualitative and quantitative data to provide a more comprehensive view of the issues~\cite{Ojanen2014}. However, the relatively smaller number of qualitative and mixed-method studies highlights a potential need to capture the nuanced social, cultural, and individual factors influencing youth online safety across Global South regions. Additionally, some articles in our dataset highlighted the need for more longitudinal studies to capture causal pathways and better understand the experiences of youth.

Our analysis also revealed that most articles made empirical (N=62) contributions, focusing on formative research aimed at understanding user behaviors, interactions, and experiences across a range of social media platforms \cite{Oloo2022Apr, Alfakeh2021May, Wang2022May}. The only other type of contribution we found within the articles were artifact-based contributions (N=4). These studies primarily involved the development of new technological solutions, focusing on gaming technology (N=3) and applications (N=1) that incorporated culturally sensitive interventions to educated and mitigate online risks \cite{cyberbullet, Reeves2024Jun}. For example, Reeves et al. \cite{Reeves2024Jun} developed a gaming system as part of a preventative education program for youth in Southeast Asia, aimed at reducing online sexual exploitation. This suggests that artifact-driven research was more experimental and design-oriented, particularly within interactive platforms like gaming, where new HCI technologies could be designed, implemented, and evaluated. 



\subsection{What: Cultural and Social Factors Influence How Regions Perceive Risks and Employ Safety Strategies} 
Our systematic review highlights six key social factors shaping youth online safety research across the Global South: 1) culture, 2) education, 3) socioeconomic status, 4) religion, 5) politics and policy, and 6) gender. These factors influenced how risks were perceived and addressed in different regions. 

\textit{Culture} emerged as the most significant social factor (N=38). In many Global South regions, cultural contexts shaped not only how risks are perceived and experienced but also how they were mitigated. Many regions in the Global South are characterized by collectivist cultures \cite{Bai2022Dec, Aljasir2022Sep, Ojanen2015Apr}, where community or family priorities often take precedence over individual concerns. Some articles revealed that this social dynamic sometimes exposed youth to online risks given the sociocultural pressures to conform to them, such as body image, sexuality, and other societal values, in fear of social exclusion \cite{Ojanen2015Apr, Hendricks2020Apr, Wang2022May}. For instance, a study by Wang et al.\cite{Wang2022May} revealed that adolescents who did not conform to societal body image ideals faced cyberbullying, being labeled unattractive by their peers.
\begin{quote}
    "\textit{Given that Saudi Arabia is a collective culture, it may be argued that the high moral discipline and level of societal commitment instilled in Saudi youth can result in lower levels of hostility in the form of cyberbullying in both genders compared to their peers in other cultures}." - \cite{Aljasir2022Sep}
\end{quote}
Culture also influenced mediation strategies employed for youth online safety \cite{Aljasir2022Sep}. Parents and adult figures played key roles in shaping youth online behaviors \cite{Reeves2024Jun, Hendricks2020Apr, Aljasir2022Sep}. Adults often relied on active mediation approaches rooted in familial and community traditions to mitigate their children's exposure to online risks \cite{wachs2021dnt, Aljasir2022Sep}. For instance, a study highlighted that the negative correlation between parent-adolescent interaction and cyberbullying is stronger in Western cultures than in most Global South regions \cite{wachs2021dnt, Ojanen2014}. Articles in our dataset attributed this to differing parenting styles; Western parenting tends to support adolescent independence, whereas parenting styles in some African, Asian, and Middle Eastern cultures are often more punitive or authoritative \cite{Aljasir2022Sep, Reeves2024Jun}. For example, a study by Aljasir and Alsebaei \cite{Aljasir2022Sep} revealed that the high moral discipline and societal commitment instilled in Saudi youth resulted in lower levels of cyberbullying.

\textit{Education} was often discussed in relation to both youth and their parents (N=20). Researchers considered education as a social factor because it underlined a significant impact on online risk levels and the effectiveness of intervention strategies. Many studies found that higher levels of education resulted in a better understanding of online risks and more effective implementation of safety strategies \cite{Teimouri2018Dec, Weru, Ming2020Mar}. Youth and parents who were more educated were better equipped to recognize and avoid risky online behaviors such as sexting, cyberbullying, or engaging in cybercrime when compared to youth or parents with less education \cite{Ming2020Mar}. However, the gap in digital literacy between parents and youth frequently emerged as a key barrier to effective online safety. Many parents, particularly in low-income or less-educated households, lacked the knowledge to effectively monitor or guide their children's online behaviors. This gap often left youth vulnerable as they navigated online spaces without proper guidance or understanding of the risks \cite{Weru, Chandrima2020Dec}. 
\begin{quote}
    "\textit{There is limited knowledge or information among the citizens in Kenya as to where their children and teenagers go online and the sites they visit. There is a general problem of generational digital divide between children, teenagers and parents as children venture into the internet where they end up sharing personal information and pictures to the unknown world}." - \cite{Weru}
\end{quote}
Despite the focus on education as a prevalent safety strategy, societal pushbacks were observed in certain regions. For instance, topics related to sexual education were often taboo, limiting the ability of schools and families to openly discuss online risks associated with sexting or sexual exploitation \cite{Id2020Nov}.

\textit{Socioeconomic status} (SES; N=13) was often discussed in connection with other social factors such as education, culture, religion, and gender. Youth from lower SES households faced significant barriers to accessing the internet and the resources needed to navigate online spaces safely \cite{Ming2020Mar}. These limitations not only impacted their understanding of online risks but also made them more vulnerable to risky behaviors, as they lacked the safety education necessary to protect themselves. 
\begin{quote}
    "\textit{With regard to befriending and meeting strangers online, the elite and urban students are significantly more aware of the danger of this practice than the rural students.}" - \cite{Ming2020Mar}
\end{quote}
Several studies highlighted that youth from lower SES backgrounds were less likely to receive formal education about online risks, leaving them more vulnerable to harmful behaviors such as cyberbullying, sexting, and even cybercrime. This knowledge gap often extended to parents, who, due to their own limited digital literacy, were unequipped to guide their children in safe online practices. A study highlighted the link between low SES and the inability to fully access or understand privacy features and online safety tools, further exacerbating the risks these youth faced online.

\textit{Religious values} were discussed (N=7) with culture as a social factor influencing youth online safety \cite{Mustafa2023Nov}. In many Global South regions, religious beliefs shaped cultural practices, which in turn influenced how online risks were perceived and addressed. For example, in conservative religious communities \cite{Id2020Nov}, certain online behaviors, such as engaging in social media or accessing specific types of content like sexual education, were viewed through a moral lens, with strict guidelines for acceptable online behavior.
\begin{quote}
    "\textit{The Confucian culture [a Religion] in Vietnam, sex and sexual education for children are considered as a taboo. Schools neglect sexual education for children, and not only parents in the countryside but also parents in the city with high levels of education also find it difficult to educate sexuality for children.}" \cite{Id2020Nov}
\end{quote}
Religious values often guided parental mediation strategies, with parents in these communities more likely to impose stricter online controls based on their beliefs \cite{Mustafa2023Nov}. While these controls were intended to protect youth, they sometimes limited open discussions about online risks, making it difficult for young people to fully understand the potential dangers they faced online.

\textit{Gender} was discussed in a few articles (N=4). Gender norms in certain regions often limit girls’ access to technology, while boys are generally afforded more freedom in their online interactions. For example, in some cultures, girls are expected to behave more obediently, which restricts their ability to fully engage with the digital world. As a result, girls in these regions may be denied access to the benefits of the internet or, conversely, be exposed to greater risks due to the societal pressures placed on them.
\begin{quote}
    "\textit{Generally, Malaysian parents have the tendency to be strict on their daughters and lax on their sons as they think that they are not at risk of being cheated or abused. This is not wise and awareness should be brought to parents that their sons can be at risk too.}" - \cite{Ming2020Mar}
\end{quote}
Some studies pointed out that gendered expectations also influenced the recruitment processes in research on youth online safety. Cultural norms around gender often dictated the types of risks boys and girls were exposed to, as well as the protective measures that were considered appropriate \cite{Ming2020Mar}. For instance, girls were more likely to be subject to stricter monitoring, while boys experienced more freedom but with fewer protective measures in place.

\textit{Politics and policy} were mentioned in only (N=1) article focusing on the role of government in setting up legislation to protect youth from online risks. This underrepresentation points to a gap in the literature, as effective government policies are critical to creating a safe online environment for youth \cite{Silva2017Privacy, Adesina2022Sep}. The single article that addressed politics and policy emphasized the need for more comprehensive laws to protect youth, particularly in regions where offline vulnerabilities, such as poverty and violence, increased their exposure to online risks. It also recommended that online safety policies be integrated into broader legislative frameworks that address both offline and online safety concerns, such as anti-cyberbullying laws and digital literacy programs in schools.


\subsection{What: Cyberbullying Most Prominent Among Youth in the Global South}
Most studies (N=44) investigated risks among five major categories: 1) Cyberbullying and online victimization, 2) privacy and data security risks, 3)sexual risks and exploitation, 4) inappropriate or harmful content, and 5) technology addiction. The remaining 20 studies used a broad, "catch-all" definition of risk, often lacking specificity regarding the exact nature of the risks \cite{Ming2020Mar, Chipangura2022COVID}. These studies generally took an exploratory approach, emphasizing the inherent dangers youth may face in online spaces without detailing specific risk types. 

Risks related to \textit{cyberbullying and victimization} emerged as the most frequently discussed (N=29). These risks spanned various aggressive online behaviors, including cyberbullying, harassment, victimization, threats, and cyberhate \cite{Chi2020Jun, Pacheco2023, Wang2022May}. A common trend within this articles was the connection between offline violence and the perpetuation of online victimization. Studies indicated that societal exposure to prejudice and bullying in physical spaces often carried over into online environments, normalizing harassment and making cyberbullying both easily perpetrated and, at times, socially tolerated \cite{Ojanen2014, Arifin2019Jun}. Additionally, these studies emphasized the severe psychological impacts of cyberbullying, including depression, anxiety, and, in extreme cases, suicide \cite{MohdFadhli2022May, sittichai2020information}. For instance, Farhangpour et al. \cite{Farhangpour2019Mar} found that cyberbullying led some South African students to contemplate suicide, a pattern echoed among Malaysian adolescents by Fadhli et al. \cite{MohdFadhli2022May}.

Risks related to \textit{privacy and data security} were also prevalent (N=14). Our findings revealed that privacy awareness among youth and their guardians in the Global South remained limited, with privacy often viewed as a secondary concern compared to the benefits of online engagement and connectivity \cite{Wardhani2021}. Usually these risks encompassed issues such as privacy violations, exploitative online data collection practices, surveillance, and financial crimes like cyberfraud and cybercrime  which often led to the unauthorized access to or misuse of youth data, leaving young users vulnerable to identity theft, targeted advertising and exploitation \cite{Silva2017Privacy, Mamat2023Apr, Adesina2022Sep}. 
Limited privacy awareness also led to emerging concerns like “sharenting,” where parents shared children’s personal content online despite fears of exposing them to long-term digital footprints. This highlighted a tension between privacy and perceived online benefits \cite{Wardhani2021, Mamat2023Apr}. Some articles pointed out the lack of youth-specific privacy policies in the Global South, which, combined with weaker regulatory frameworks, left young users more vulnerable. For example, Silva et al. \cite{Silva2017Privacy} found that Western-centric privacy laws failed to meet the specific needs of Brazilian youth.
Additionally, a few studies (N=2) linked privacy risks like financial scams and cybercrime to broader socio-economic factors, revealing youth as both enablers and victims of these issues. Adesina et al. \cite{Adesina2022Sep} noted economic hardship as a driver of cybercrime involvement among Nigerian youth, underscoring the importance of addressing underlying socio-economic conditions to effectively mitigate privacy risks.

There were some articles focused on \textit{sexual risks} (N=8),including sexting, cybergrooming, pornography, and sex trafficking \cite{Reeves2024Jun, Ghorashi2019Feb, RocioGrooming, Id2020Nov}. Sexting seemed to be the most commonly studied sexual risk. Our findings revealed that while some youth willingly engaged in risky sexual behaviors, such as sexting or viewing pornography, they often lacked the awareness to fully understand potential consequences. Others faced harassment for declining participation in these behaviors \cite{Pacheco2023}. Hence, researchers emphasized the need of educating youth and their guardians about online sexual risks, including potential consequences, and equipping youth with strategies to better recognize and mitigate these risks. \citet{RocioGrooming}, for example, explored the relationship between emotional intelligence and cybergrooming among Mexican youth. Their study found that higher emotional intelligence reduced vulnerability to cybergrooming, highlighting the importance of educational interventions that foster emotional intelligence to help youth navigate online risks more effectively. Despite these efforts, several articles highlighted that cultural and religious sensitivities around sexual risks and education created barriers in some regions of the Global South, making it challenging for researchers to collect data or address these topics directly. These sensitivities can also restricted schools and community programs from incorporating sexual education into their curricula, reducing opportunities for awareness and support.
\begin{quote}
   "\textit{Currently, sexual education for  children  in  Vietnam  does not meet  the needs  of children  to  understand  this issue because  in  the  Confucian culture  in  Vietnam, sex  and  sexual  education  for  children  are considered as a taboo}" - \cite{Id2020Nov}
\end{quote}

Finally, only (N=3) studies explored \textit{technology addiction} \cite{Tech_Addiction}, while (N=2) studies specifically examined youth exposure to \textit{inappropriate or harmful content}, such as online drug and alcohol marketing\cite{Bai2022Dec}. These articles highlighted the need to examine how prolonged exposure to technology affects youth well-being with more emphasis on providing age-appropriate content\cite{Tech_Addiction}.\\


\subsection{What: Promoting Safety Through Education and Awareness}
We identified several key strategies employed in the Global South to enhance youth online safety. We categorized these strategies into five main themes: 1) education and awareness, 2) mediation and support systems, 3) privacy and security, 4) resilience and emotional well-being, and 5) policy and regulation.

\textit{Education} emerged as the most extensively discussed safety strategy (N=41). These studies emphasized the importance of educating youth and primary stakeholders within their social ecosystems (including parents, teachers, and community leaders) about online risks and safer online practices. The majority of these articles advocated for access to tailored educational interventions that consider the unique social and cultural contexts of youth in the Global South, focusing on areas such as adolescent education, privacy programs, sex education, violence prevention, and digital literacy initiatives \cite{Popovac2020Dec, sittichai2020information}. While others proposed educational tools as interventions to promote online safety education \cite{PerezCyberkids}.

A recurring theme within articles studying this strategy was the recognition of youth autonomy. Hence, many studies \cite{cyberbullet, MohdFadhli2022May, BlancaflorPrivacy, Popovac2020Dec} highlighted the importance of empowering young people by not only informing them about potential online risks but also helping them understand the consequences of their online actions and equipping them with effective risk mitigation strategies. Additionally, the importance of stakeholder involvement, particularly the roles of parents, teachers, and community figures, was frequently mentioned 
\cite{Weru, Choudhury2023Apr, BlessingNamibia}. In many Global South contexts, youth rely on these figures for guidance, and the studies underscored the need for educational resources and technical literacy that help parents and adult stakeholders better understand online risks and current safety strategies to effectively mediate youth's online experiences. For instance, Alfakeh et al.’s study on parental perceptions of cyberbullying in Saudi Arabia demonstrated that proactive parental engagement in addressing online bullying can significantly reduce its occurrence, stressing the need for digital literacy programs aimed at educating parents \cite{Alfakeh2021May}.
\begin{quote}
    "\textit{There is limited knowledge or information among the citizens in Kenya as to where their children and teenagers go online and the sites they visit.}" - \cite{Weru}
\end{quote}
A smaller subset of articles (N=4) explored innovative approaches, such as gamified educational interventions. These approaches incorporated culturally relevant, interactive methods to engage young users and promote safer online behaviors \cite{PerezCyberkids, Reeves2024Jun}. For example, Perez et al. developed a serious game designed to teach youth aged 8 to 12 about cybersecurity, offering them a playful yet educational way to grasp the responsibilities tied to using digital technologies. These interactive, gamified methods were found to be particularly effective in fostering engagement and improving online safety practices among young users. Overall, articles in this category of research highlighted a consensus on the need for contextually sensitive education and awareness programs that empower youth and provide them with the knowledge and tools necessary to navigate their online experiences safely.

\textit{Mediation and support systems} emerged as the second most common online safety strategy in the Global South (N=21). This strategy emphasized the importance of parental, and stakeholder involvement in helping young users navigate their online experiences. A notable theme across these studies was the tension between overly restrictive mediation and more active, supportive mediation strategies. While restrictive approaches such as strict monitoring or limiting access to technological devices were commonly employed \cite{Reginasari2021, Chen2023}, many studies highlighted their ineffectiveness in improving youth safety. Instead, restrictive measures often led young users to circumvent rules, thereby increasing their exposure to risks rather than reducing them \cite{Choudhury2023Apr}. These restrictive strategies were mainly employed in schools by teachers when mediating online technology use by youth \cite{Chipangura2022COVID}. 

Studies by Wachs \cite{wachs2021dnt} and Chen \cite{Chen2023}, for instance, stressed that active mediation, grounded in open communication and trust between parents and children, was more effective in supporting youth online safety. Active mediation, according to papers in our dataset, entailed parents guiding, monitoring, and supporting their children's online activities without overly controlling them, fostering a relationship where youth feel comfortable sharing their online experiences. This approach helped build trust, which in turn empowered youth to make safer choices online \cite{Chen2023, Reginasari2021}. For instance, a study by Chen et al. \cite{Chen2023} found that teenagers whose parents practiced high active mediation combined with low restrictive mediation were less likely to experience cyberbullying. Their findings highlighted the importance of balanced mediation, where parents remain engaged and supportive but not overly controlling. 

In addition to parental involvement, community norms, and values were shown to play a significant role in mediating youth online behaviors, particularly in collectivist societies common across the Global South. Cultural expectations and societal standards often shaped and dictated how youth engaged with social media \cite{Gohal2023Dec}. For example, a study by Priya et al. \cite{priya2023} revealed that in India, societal norms dictate stricter expectations for girls' online behavior compared to boys. Girls are expected to act more obediently, and any perceived misbehavior, whether in real life or on social media, is seen as a dishonor to the family. This pressure creates gendered differences in how youth experience and navigate online spaces.
\begin{quote}
    “\textit{In Indian families, girls are expected to behave more obediently than boys; this is also regarded as a social norm. As a result, any unruly behavior by girls in society or on social media is strongly viewed as a discredit to families, [...] they are concerned that their freedom of access and online presence will be limited when their parents see them.}” - \cite{priya2023}
\end{quote}

Overall, the studies underscored the importance of active, culturally sensitive mediation, highlighting that parental involvement, community values, and trust-based approaches are critical in shaping safer online behaviors for youth in the Global South.

\textit{Privacy and security strategies} were also discussed (N=6), with a focus on employing technical solutions such as privacy features, AI-driven fraud detection, online moderation, and URL filtering aimed at improving the cybersecurity of youth \cite{Silva2017Privacy, KavikairiuaALGO}. While these measures are important for safeguarding youth online, they are less commonly explored compared to other safety strategies. These studies often highlighted the need for such technical solutions to help both youth and adult stakeholders prevent or mitigate risky online experiences.

For example, a study by Hendricks et al. \cite{Hendricks2020Lgbtq} revealed that many LGBTQIA+ youth resort to privacy-related measures, such as blocking users, deleting offensive content, and adjusting privacy settings, as coping mechanisms for dealing with cyber-victimization. This proactive use of privacy tools illustrated how youth actively engaged with available security features to protect themselves in hostile online environments. Similarly, another study by Silva et al. \cite{Silva2017Privacy} examined the usability of Facebook’s privacy settings and their impact on the safety of Brazilian children and teenagers. Their findings indicated that usability challenges hindered many young users from fully utilizing these security and privacy features as many Brazilian youth had limited knowledge of Facebook’s privacy settings, which ultimately restricted their ability to protect themselves effectively.

These studies collectively pointed to the need for user-friendly and intuitive privacy solutions that empower youth and other stakeholders to manage their online safety. Privacy features, while vital, required better integration with educational initiatives to ensure that young users were both aware of and capable of leveraging them to protect their privacy and security in online environments \cite{cyberbullet}.

Emotional intelligence and resilience were only found in a small subset of studies (N=6). These studies often stressed the importance of fostering mental and emotional support systems that empower youth to face online challenges, recognizing that resilience-building is essential for navigating the inherently risky nature of digital environments. For example, a study by Gan et al. \cite{Gan2022} found that teens with higher levels of emotional intelligence were significantly less likely to experience cyberbullying compared to their peers with lower emotional intelligence, particularly in situations where students felt alienated from school. This suggests that emotional intelligence served as a protective factor, enabling youth to manage and mitigate harmful online interactions.

Moreover, the role of cultural practices in fostering emotional resilience was also explored in some studies within this safety strategy. For instance, Wang et al.'s study \cite{Wang2022May} examined how certain cultural practices helped Chinese youth build the emotional resilience needed to moderate their engagement in cyberharassment. Their findings underscored the importance of tailoring emotional support systems to specific cultural contexts, recognizing that resilience-building strategies may vary across different regions and communities. Overall, Emotional Intelligence and Resilience strategies entailed the development of mental and emotional support systems for youth, which can strengthen their ability to withstand and bounce back from online risks. This approach complements more technical safety strategies by focusing on the internal capacities that youth can develop to protect themselves.

\textit{Policy and regulation} aimed at mitigating online risks for youth were the least frequently discussed safety strategy (N=5). These articles primarily focused on regulatory measures and restrictive strategies designed to protect young people from online harm. Despite being less explored compared to strategies like education or mediation, studies within this safety category emphasized the essential role of governance and protective legislation as part of a comprehensive approach to youth online safety. A recurring theme across most of the articles was the need for an authoritative legal framework to hold perpetrators accountable for online risks such as cyberbullying and the inappropriate marketing of substances to youth \cite{Kim2021Feb, Farhangpour2019Mar}. 

Additionally, some articles also emphasized the importance of policy in addressing the broader social and economic factors contributing to youth vulnerability online. For instance, Adesina’s study \cite{Adesina2022Sep} advocated for policies that empower youth by providing employment opportunities, which could, in turn, reduce the prevalence of cybercrime by tackling the root causes of risky online behaviors.

\section{Discussion}
In this section, we discuss the implications of our findings for HCI researchers, offering insights into potential applications and areas of focus. Finally, we address the limitations of our study and suggest directions for future research that can address these gaps.

\subsection{Moving Beyond a One-Size-Fits-All Approach for Youth Online Safety in the Global South (RQ1 and RQ2)}
Findings from our systematic review revealed that youth online safety research in the Global South is developing rapidly, with a strong emphasis on formative research aimed at exploring, understanding, and describing the nature of online risks and behaviors among youth \cite{Rivera2016, Arifin2019Jun}. Most studies adopted empirical approaches, examining these issues through sociocultural lenses such as collectivist values, gender norms, and religious and cultural sensitivities. For instance, in many Global South regions, there is a greater emphasis on community, family involvement, and collective responsibility, which can significantly shape youth online behavior and perceptions of safety. While the findings in our review highlighted a limited number of concrete intervention strategies, they suggested that researchers were increasingly aware of the need to develop culturally tailored solutions to mitigate risks \cite{gair2024risk, herkanaidu2021towards}. One of the positive trends observed was the growing interest among researchers affiliated with institutions in the Global South in exploring youth online safety\edit{; this reflects a regional awareness of the importance of addressing these issues locally \cite{Wang2022May, Chandrima2020Dec, MohdFadhli2022May}. Yet, we found a geographic imbalance in the reviewed studies, with a significant number of online safety papers from Asia and fewer from Africa and Latin America. This may result from systemic inequities in research, such as funding and publication practices as well as historic sociopolitical challenges. In many Global South regions, research is primarily funded by non-profits, international agencies, and philanthropic foundations \cite{Chankseliani2023Jan}. However, these funding sources often lack support for overhead costs necessary to strengthen institutional research capacity. Additionally, financial barriers, such as high publication fees, limit researchers' abilities to publish in major academic journals, leading to underrepresentation in globally visible literature \cite{Naidu2024Apr}. Addressing this imbalance requires equitable funding distribution, sustainable support for research infrastructure, and diverse publication avenues to ensure studies from underrepresented regions are accessible and acknowledged.} 

Overall, our findings highlight a critical need for HCI research to adopt a more holistic and globally inclusive perspective when investigating youth online safety. The inclusion of cross-regional studies with Global North collaborators indicated an emerging interest in global partnerships \cite{wachs2020routine}, which could facilitate the sharing of best practices in addressing youth online safety across both regions. We call for such cross-cultural partnerships that can bridge knowledge gaps and foster more comprehensive and contextually relevant solutions for youth online safety. By incorporating insights from the Global South, \edit{cross-regional collaborations can develop a nuanced understanding of how cultural, social, and economic factors influence youth online experiences. This approach not only enriches the research landscape but also promotes the evaluation of effective, context-specific interventions \cite{Reynolds2023bridges}.} Additionally, socioeconomic factors, such as varying levels of access to technology and differing levels of digital literacy, can influence how youth encounter and manage risks online. A one-size-fits-all approach, rooted in assumptions from the Global North, may overlook these complex realities, leading to interventions that fail to engage youth effectively. A more inclusive approach would mean that safety interventions are designed with these cultural differences in mind, making them more relevant and effective for youth across diverse regions. This might involve adapting digital tools to accommodate local norms around privacy or creating educational content that aligns with regional attitudes towards authority and family roles.

\subsection{Methodological Best Practices and Paths Forward for Online Safety Research in the Global South}
Most of the Global South online safety studies employed survey methods to gather quantitative data to identify the prevalence and patterns of online risks among youth. This approach was typical of the early stages of research in new fields, where establishing a baseline understanding was crucial. \edit{However, relying solely on quantitative research limits the opportunities to capture contextual insights and subjective experiences, such as changes in emotions and behaviors of respondents \cite{Queiros2017Sep}.} Global literature have indicated that qualitative methods—such as interviews, focus groups, and ethnographic studies were essential for capturing the deeper, contextual nuances of youth experiences \edit{\cite{badilfoster, pam2016deardiary}.} In the culturally diverse settings of the Global South, qualitative research could have provided richer insights into how young people perceived and responded to online risks and how sociocultural dynamics shaped their digital behaviors. Integrating qualitative approaches could have offered a more comprehensive understanding of the complex interplay between cultural norms and online safety practices.

Our findings indicate that youth online safety research in the Global South often adopted a quantitative study design that employs surveys and statistical analyses. While these methods provided valuable data on the prevalence and patterns of online behaviors, they did not fully capture the complex social and cultural dynamics that influence youth behavior online. For instance, a survey might reveal high rates of cyberbullying but may not explain why certain forms of bullying are more prevalent or how cultural norms around conflict resolution influence these behaviors. To address these gaps, HCI scholars should prioritize mixed-methods research \cite{hardy2019rural, pinter2017adolescent, anuyah2023cultural}, combining quantitative data with in-depth qualitative approaches such as interviews, focus groups, and ethnographic studies. This would allow researchers to gain a deeper understanding of the contexts in which risks occur and how youth navigate them. Additionally, the HCI community has long emphasized the role of participatory design in ensuring the inclusion of marginalized communities \cite{anuyah2023cultural, agha2023strike, badillo2021conducting}; our findings further underscore the critical importance of involving young people directly in the creation of online safety interventions \cite{cyberbullet}. Engaging youth, particularly those from the Global South, through co-design allows for developing tools and strategies that genuinely reflect their needs and preferences. This participatory approach not only amplifies youth voices but also ensures that digital safety tools are more relevant, culturally attuned, and effective. By involving youth in the design process, their agency is recognized, and valuable insights are gained into how they perceive risks and navigate online environments \cite{heidi2016}. \edit{However, implementing participatory methods in low-resource settings presents challenges such as language barriers, literacy constraints, power imbalance, and limited technology access \cite{harrington2019deconstructing, sharifa}. To address these barriers, researchers can employ various engagement strategies, such as visual storytelling and offline participatory workshops that do not rely on digital connectivity~\cite{maslow}. Collaborating with local educators, community leaders, and youth mentors can help bridge accessibility gaps and ensure meaningful participation \cite{Oguine2024CHINS, mun_education}. At the same time, we need to acknowledge that participatory design may not always be democratic or equitable, and may even be harmful, when working with communities across the world, where norms of speaking up, power, and oppression differ from the Global North, and play a critical role in what participation looks like \cite{sharifa, wani2022hartal}.}

 \subsection{Attention to Socio-Cultural Norms in Researching Youth Online Safety}
Our findings underscored the significant role of sociocultural norms in shaping the risks youth face online and the strategies employed to ensure their safety. This was consistent with broader literature emphasizing the importance of understanding online behaviors within the context of offline social and cultural environments \cite{backe2018networked, Ojanen2014}.
Researchers in the Global South were deliberate in emphasizing the socio-cultural contexts of their study areas. They frequently highlighted how cultural norms influenced youth interactions online. This reflected an awareness that youth in the Global South navigated digital spaces differently than their peers in the Global North, where individualistic values often prevailed. For example, in many Asian and African contexts, cultural norms such as respect for elders and community cohesion \cite{bell2011confucianism, yi2024reinforcing} led to a preference for parental or adult mediation in managing youth online experiences. This was consistent with broader literature suggesting that in collectivist societies, familial and community guidance played a critical role in shaping youth behavior both online and offline \cite{chandni, Alfakeh2021May}. However, our review also noted potential challenges with this approach, such as the risk of overly restrictive mediation. Such restrictions could prompt youth to seek out digital platforms covertly, inadvertently increasing their exposure to online risks \cite{pam2017parents, Wisniewski2015Feb}.
While many studies in our dataset focused on gender differences between male and female youth \cite{Ojanen2014}, there was a significant lack of research on marginalized groups, including LGBTQIA+ youth \cite{Hendricks2020Apr}, youth with disabilities \cite{Leung2023Autism}, and those from minority ethnic communities. This mirrored a broader gap in online safety literature \cite{oguine2023you, badillo2017abandoned, Oguine2024CHINS}, where studies often overlooked the unique experiences and needs of marginalized youth. For instance, LGBTQIA+ youth might face distinct risks related to identity disclosure, while youth with disabilities might encounter accessibility challenges when using privacy settings or reporting tools. Addressing this gap was essential for developing inclusive safety strategies that catered to the diverse realities of all young users.
Our review found that the nature of online risks, such as cyberbullying, often mirrored offline discrimination and violence. This finding aligned with broader literature, which recognized that digital environments frequently reflected and even amplified existing societal tensions. Many studies in our dataset also noted that discussions around sexual risks were less common, likely due to cultural taboos surrounding such topics in some Global South regions \cite{Id2020Nov}. \edit{These cultural norms can hinder the implementation of sexual health interventions by limiting open dialogue and access to information, thereby perpetuating misinformation and stigma \cite{Tohit2024Aug}.} This underscores the necessity for culturally sensitive research approaches that address these risks without neglecting them due to societal discomfort. Effectively tackling these sensitive issues requires a nuanced strategy that respects cultural norms while prioritizing the safety and well-being of youth.

\subsection{Implications of the Findings on Youth Online Safety (RQ3)}
The evolving landscape of youth online safety in the Global South reveals critical opportunities for expanding the scope of current research and intervention strategies. By considering the unique cultural, social, and economic factors that shape youth experiences in these regions, HCI researchers can move towards more inclusive and impactful approaches to youth online safety. Below, we discuss some of the implications of our study.

\begin{itemize}
    \item \textbf{Recognizing Resourcefulness and Resilience:}
    An important implication of the findings is the need to shift towards strength-based approaches in youth online safety research. Rather than focusing solely on deficits or challenges faced by youth in the Global South, HCI researchers should identify and emphasize the resourcefulness, resilience, and adaptive strategies that young people and their communities already possess \cite{hardy2019rural}. Many youths in these regions develop creative solutions to mitigate online risks, often leveraging community support, peer networks, and culturally relevant practices. \edit{For instance, in areas with limited access to formal education, young people have established community-based learning hubs that utilize locally available materials and peer-to-peer teaching methods to enhance online safety awareness~\cite{Wang2022May, Alfakeh2021May}.} Highlighting these strengths can provide a more balanced understanding of youth online experiences and offer insights into community-driven strategies that could be adapted or scaled in other contexts. By adopting a strength-based perspective, researchers can ensure that their work respects and builds upon the existing capacities within these communities, fostering empowerment and sustainability in online safety interventions.

    \item \textbf{Designing Contextually Adaptive Tools and Interventions:}
     Our findings highlight the importance of designing educational interventions and tools that capture the socio-cultural landscape of the Global South \cite{reeves2024May, cyberbullet, PerezCyberkids}. This underscores the need for HCI researchers and designers to develop tools, platforms, and policies that are adaptable to local norms and cultural practices. Rather than applying one-size-fits-all solutions, there is a growing demand for customizable privacy features and safety settings that align with regional expectations, such as varying degrees of parental involvement or community-based norms around privacy \cite{Silva2017Privacy}. For instance, developing parental control programs that incorporate multilingual options to help overcome language and literacy barriers often faced by non-English-speaking or less educated parents \cite{Ndalaye2022Oct}. These adaptations could make educational resources and privacy features more inclusive and effective for diverse communities across the Global South.
    
    \item \textbf{Enhancing Cross-Regional Collaboration:}
    Cross-regional collaboration between researchers from the Global North and Global South is crucial for advancing a comprehensive understanding of youth online safety on a global scale. Findings from our systematic review highlighted the importance of such collaborations in bridging knowledge gaps, enabling a richer understanding of the diverse risks and protective strategies employed across different regions \cite{wachs2020routine, Rivera2016, WachsHITem, DellHCI4D}. This type of collaboration enables a dynamic exchange of expertise, \edit{where researchers from both the Global South and Global North} contribute valuable insights into socio-cultural factors shaping online behaviors, including community norms, familial dynamics, and region-specific risks \cite{Oloo2022Apr, wachs2020routine}. \edit{At the same time, researchers across regions can share diverse methodological approaches, technological innovations, and policy strategies, ensuring that interventions are both contextually relevant and globally informed.} This partnership helps bridge knowledge gaps and fosters more comprehensive and contextually relevant solutions for youth online safety. Also, this collaboration can facilitate a comparative evaluation and understanding of how different regions approach youth online safety, helping identify best practices adaptable across cultural contexts.

\end{itemize}

\subsection{Limitations and Future Research}
Despite the thoroughness of our systematic review, there are several limitations that need to be acknowledged. First, there was a notable geographic imbalance, with South America and Africa underrepresented and no representation from the Caribbean or Oceania. This likely reflects broader trends in research funding and infrastructure. We encourage future studies to expand the geographic scope of youth online safety research to better represent these regions. Additionally, despite efforts to include diverse languages, only four articles in our dataset were in Spanish, with most in English, suggesting that studies in other languages may have been overlooked. This highlights the need for more inclusive research practices, including translating non-English literature to capture diverse perspectives. Finally, although our review examined a variety of online risks, there was a noticeable focus on traditional concerns such as cyberbullying and privacy. This emphasis may have limited attention to emerging risks, particularly those related to AI-driven content recommendation and generative AI. Future research should explore these evolving challenges to provide a more comprehensive understanding of the online landscape and its implications for youth.

\section{Conclusion}
Our paper captures the emerging landscape of youth online safety research in the Global South, assessing current approaches and identifying future directions. Through a systematic review of existing studies, we highlight the critical role of cultural factors in shaping online risks and interventions, revealing a predominant focus on education and awareness as foundational safety strategies. Our findings underscore the need for methodological diversity and the inclusion of youth voices to better capture the sociocultural contexts that influence online behaviors. We also point out gaps in geographic representation and the need for more culturally attuned approaches to online safety. Finally, we outline the contributions of this paper to researchers, educators, and technology designers interested on youth online safety, aiming to inspire future work that addresses the unique needs of young people in the Global South and promotes a safer, more inclusive digital world.

\begin{acks}
This work is supported in part by the ND-IBM Funding Award. Any opinions, findings, conclusions, or recommendations expressed in this material are those of the authors and do not necessarily reflect the views of our sponsors.
\end{acks}



\bibliographystyle{ACM-Reference-Format}
\bibliography{Sections/References}

\newpage
\appendix
\section{Summary Codebook}
\begin{table}[h]
\caption{Summary Codebook}
\label{tab:codebook}
\centering
\scriptsize
\begin{tabular}{p{1.2cm}|p{4cm}|>{\raggedright\arraybackslash}p{5cm}|>{\raggedright\arraybackslash}p{1.7cm}} 
\hline
\rowcolor[HTML]{EFEFEF}
\textbf{Dimensions} & \textbf{Categories (Short Descriptions)} & \textbf{Codes (No. of Articles)} & \textbf{Examples} \\ \hline

\multirow{4}{*}{\textbf{Where?}} 
 & \textbf{Geographical location }\newline \textit{(Region that was studied)} & Asia (48) & \cite{Ming2020Mar, Alfakeh2021May, Bai2022Dec, Mustafa2023Nov, Leung2023Autism} \\[-3ex] \cline{3-4}
 &  & Africa (12) & \cite{Farhangpour2019Mar, Chipangura2022COVID, BlessingNamibia} \\[0ex] \cline{3-4}
 &  & South America (8) & \cite{RocioGrooming, Rivera2016, Silva2017Privacy} \\[0ex] \cline{2-4} 
 & \textbf{Authors' affiliation} \newline \textit{(Country where authors are from)} & Eritrea, Kenya, Namibia, Nigeria, South Africa, The Gambia, China, Brazil, Chile, Colombia, Mexico, Uruguay, Cambodia, Indonesia, Malaysia, Philippines, Taiwan, Thailand, Vietnam, Bangladesh, India, Iran, Nepal, Pakistan, Sri Lanka, Saudi Arabia, Syria, Cyprus, Germany, Greece, Singapore, South Korea, Spain, USA &  \\ \hline\hline
\multirow{6}{*}{\textbf{Who?}} 
 & \textbf{Population studied} \newline \textit{(Type of group examined)} & Typical youth (65) & \cite{Mustafa2023Nov, Bai2022Dec, Farhangpour2019Mar, Ojanen2014} \\[-3ex] \cline{3-4}
 &  & Youth with disability (1) & \cite{Leung2023Autism}\\ \cline{3-4}
 &  & LGBTQIA+ (1) & \cite{Hendricks2020Lgbtq} \\ \cline{2-4} 
 & \textbf{Study participants} \newline \textit{(Stakeholders that participated in study)} & Youth (54) & \cite{Hendricks2020Apr, PerezCyberkids, Mustafa2023Nov, Ojanen2014} \\[-3ex] \cline{3-4}
 &  & Parents (10) & \cite{Wardhani2021, Chen2023} \\[0ex] \cline{3-4}
 &  & Teachers (1) \&  Counselor (1) & \cite{Chipangura2022COVID, Choudhury2023Apr} \\ \hline\hline
\multirow{6}{*}{\textbf{How?}} 
 &\textbf{Research design} \newline \textit{(Overall strategy to answer research questions)} & Quantitative (46) & \cite{Farhangpour2019Mar, Silva2017Privacy, BlessingNamibia, Wang2022May} \\[-3ex] \cline{3-4}
 &  & Qualitative (17) & \cite{Mustafa2023Nov, Ojanen2015Apr, Wardhani2021}\\[0ex] \cline{3-4}
 &  & Mixed-method (3) & \cite{Iqbal2021Dec} \\[0ex] \cline{2-4}
 & \textbf{Study method} \newline \textit{(Method of data collection)} & Survey (52) & \cite{Farhangpour2019Mar, BlessingNamibia, Gan2022, Wang2022May} \\[-3ex] \cline{3-4}
 &  & Interview \& Focus group (16) & \cite{Mustafa2023Nov, Iqbal2021Dec, Ojanen2015Apr}\\[0ex] \cline{3-4}
 &  & User study (2) & \cite{PerezCyberkids} \\[0ex] \cline{3-4}
 &  & Dataset (1) & \cite{Andrews} \\[0ex] \cline{3-4}
 &  & Case study (1) & \cite{chandni} \\[0ex] \cline{2-4}
 & \textbf{HCI Contributions}  \newline \textit{(Based on Wobbrock and Kientz \cite{wobbrock})} & Empirical (62) &  \cite{Farhangpour2019Mar, BlessingNamibia, Gan2022, Wang2022May}\\[-2ex] \cline{3-4}
 &  & Artifact (4) & \cite{reeves2024May, Quayyum2024Jun} \\ [1ex] \hline\hline

\multirow{5}{*}{\textbf{What?}} 
 & \textbf{Cultural and Societal Influences} \newline \textit{(Measures studied related to values, beliefs, and norms)}  & Culture (38) & \cite{Mustafa2023Nov, Hendricks2020Lgbtq, Ojanen2014, MohdFadhli2022May} \\[-5ex] \cline{3-4}
 &  & Education (20) & \cite{BlessingNamibia, Arifin2019Jun, Vieira2022May} \\[0ex] \cline{3-4}
 &  & Socio-economic status (13) & \cite{Pacheco2023, Priya2024May} \\[0ex] \cline{3-4}
 &  & Religion (7) & \cite{Mustafa2023Nov, Iqbal2021Dec} \\[0ex] \cline{3-4}
 &  & Gender (4) & \cite{MohdFadhli2022May, Ming2020Mar} \\[0ex] \cline{3-4}
 &  & Politics and Policy (1) & \cite{Adesina2022Sep} \\[0ex] \cline{2-4}
 & \textbf{Online Risks} \newline \textit{(Types of online risks studied)} & Cyberbullying \& Online Victimization (29) & \cite{Ojanen2015Apr, Farhangpour2019Mar, Alfakeh2021May, Wang2022May} \\[-2ex] \cline{3-4}
 &  & Privacy \& Data Security Risks (14) & \cite{Adesina2022Sep, Silva2017Privacy}\\[0ex] \cline{3-4}
 &  & Sexual Risks \& Exploitation (8) & \cite{Id2020Nov, RocioGrooming} \\[0ex] \cline{3-4}
 &  & Inappropriate or Harmful Content (2) & \cite{Bai2022Dec} \\[0ex] \cline{3-4}
 &  & Technology Addiction (3) & \cite{Chandrima2020Dec}\\[0ex] \cline{2-4}
 & \textbf{Safety Strategies} \newline \textit{(Recommended strategies for promoting online safety)} & Education and Awareness (41) & \cite{Oloo2022Apr, Choudhury2023Apr, Wardhani2021, Popovac2020Dec} \\[-4ex] \cline{3-4}
 &  & Mediation and Support Systems (21) & \cite{Priya2024May, Gohal2023Dec, Chen2023} \\[0ex] \cline{3-4}
 &  & Privacy and Security (6) & \cite{Silva2017Privacy} \\[0ex] \cline{3-4}
 &  & Emotional Resilience (6) & \cite{RocioGrooming} \\[0ex] \cline{3-4}
 &  & Policy and Regulation (5) & \cite{Alfakeh2021May}\\ \hline

\end{tabular}
\end{table}

\end{document}